\documentstyle[12pt]{article}
\newcommand{\be}{\begin{equation}}
\newcommand{\ee}{\end{equation}}

\begin{document}
\pagestyle{empty}
\title{On the Analytic Structure\\ of the Quark Self-Energy\\
in Nambu--Jona-Lasinio Models}
\author{P. P. Domitrovich and H.\ M\"{u}ther
\\\\
Institut f\"{u}r Theoretische Physik,\\ Universit\"{a}t T\"{u}bingen,\\
D-72076 T\"{u}bingen, Germany}
\maketitle

\date{\today}

\begin{abstract}
The self-energy of quarks is investigated for various models which are
inspired by the Nambu--Jona-Lasinio (NJL) model. Including, beyond the
Hartree-Fock approximation, terms up to second-order in the quark
interaction, the real and imaginary parts
of scalar and vector components of the self-energy are discussed. The
second-order contributions depend on the energy and momentum of the quark
under consideration. This leads to solutions of the Dirac equation
which are significantly different from those of a free quark or a quark
with constant effective mass, as obtained in the Hartree-Fock
approximation.
\end{abstract}
\pagestyle{empty}

%\pacs{}

\clearpage
\pagestyle{headings}

\section{Introduction}

For low momentum transfer between quarks, where asymptotic freedom of
QCD is no longer applicable, the Nambu--Jona-Lasinio (NJL) model \cite{njl},
reinterpreted for quarks, has received much attention during the last
few years \cite{klev}. This model utilizes an effective two-body
quark-quark interaction and concomitantly displays
several of the essential features of QCD. For example, the interaction
part of the NJL Lagrangian exhibits chiral symmetry, and for interaction
strengths beyond a critical value this symmetry is broken dynamically.
The breaking of chiral symmetry, which is
obtained in the mean-field approximation, yields a sizable quark-antiquark
pair condensation and leads to constituent quark masses which are much
larger than the bare current quark masses considered in the Lagrangian.
Although confinement is not included in the NJL model,
this model has been successful in accounting for important features of
QCD  at low energies and can be considered
as a low-energy approximation to QCD \cite{alkof}.

In most cases, the NJL model is treated in the mean-field
or Hartree-Fock approximation in order to study quark properties.
As already mentioned above, a large scalar contribution to the
self-energy of the quarks results and gives rise to an effective mass. It is
easy to adjust the parameters of the model, strength of the
interaction, and cutoff parameter in such a way that the effective
masses are close to the empirical masses of the constituent quarks
\cite{ulf,hats,hen}. As a next step one can study the properties
of the mesons as predicted by this model. For that purpose one uses the
single-quark Green's function, as determined in the Hartree-Fock
approximation, and evaluates the polarization propagator or response
function for the various quark - antiquark ($q \bar{q}$) channels within
the RPA approximation. The residual interaction yields very collective
excitation modes, in particular, for the isovector-pseudoscalar and the
scalar-isoscalar excitations, and the energies of these collective
modes are associated with the masses of the corresponding mesons ($\pi$
and $\sigma$ mesons).

As the quark interaction of the NJL model gives rise to such collective
$q \bar{q}$ excitation modes, it seems quite natural to go beyond the
mean-field approximation and study the effects of the interaction of
quarks with the $q \bar{q}$ excitations on the self-energy of
quarks. One approach in this direction is to consider the effects
arising from the $\pi$-exchange term \cite{cao}. Another way is to
evaluate all contributions of second order in the interaction (see
Figure 1) using standard many-body theory techniques \cite{serot,chin}.
These two approaches supplement each other in the sense that the
former one only accounts for one special collective $q \bar{q}$ mode,
whereas the second approach considers all $q \bar{q}$ excitations,
ignoring the collective features which are due to the residual
interaction.

The effects of the second-order terms on the real, scalar part of the
self-energy have been investigated in a previous paper \cite{domi}.
These contributions depend on the energy and momentum of the quark
under consideration, and it was found that, depending on the details of
the Lagrangian under consideration, the effects can be a sizable
correction of the Hartree-Fock contribution. In contrast to the
Hartree-Fock approximation, however, the second-order terms also yield
a vector component for the self-energy and imaginary parts for both the
scalar and vector components of the self-energy.

The terms displayed in Figure 1 describe the admixture of $2q-1\bar{q}$
and $2\bar{q}-1q$ components to the propagation of the quarks and
antiquarks, respectively. Therefore, after one has determined the
complete analytic structure of the quark self-energy on this level, one
can determine the corresponding Green's function and evaluate the
$q \bar{q}$ response function. In this way one accounts for the
admixture of $2q-2\bar{q}$ configurations to the polarization
propagator, which should provide a more realistic description, in
particular, for the mesons like the $\sigma$, which contain large $2\pi$
contributions. Therefore, the present investigation can be understood
as a step towards a consistent treatment of the NJL model beyond the
mean-field approximation (see also discussion in \cite{domi}).

As in our earlier paper \cite{domi}, four Lagrangians
will be considered, adjusting the coupling strength in such a way that
the same constituent mass is obtained for quarks of zero momentum in
each model. In Section 2, a brief description of the four Lagrangians
is presented, while in Section 3, the formulas for the scalar and vector
self-energies are discussed.
Section 4 contains the results of the calculations accompanied by a
discussion of these results, and the main
conclusions are summarized in Section 5.

\section{Lagrangians motivated by the NJL model}

In its original form \cite{njl}, the NJL model was designed to describe
a system of interacting nucleons, but has more recently been reinterpreted
as a quark Lagrangian of identical form:
\be
{\cal L}_{D}^{(A)}={\bar\psi}(i\nabla\hskip-0.8em{/}-m_0)\psi
+G_{A}\left[(\bar\psi\psi)^2
+(\bar\psi{i}{\gamma}^{5}\vec\tau\psi)^2\right], \label{eq:njla1}
\ee
such that the Fermion field $\psi$ now represents
quarks with SU(2) flavor ($\vec\tau$ denoting the Pauli matrices for
the flavor degrees of freedom) and SU(3) color.
For vanishing current
quark mass $m_{0}$, this Lagrangian is invariant under a chiral
transformation.
Particular to this study, A denotes the first of the four models considered
and the subscript $D$
implies that only the direct part of the contact interaction defined in
eq.(\ref{eq:njla1}) is included.

For such a contact interaction, the Pauli-exchange terms originating from
the Lagrangian in eq.(\ref{eq:njla1}) are easily obtained by performing
a Fierz transformation (in Dirac, flavor, and color spaces) of the
interaction term
\be
G_{A}\left[(\bar\psi\psi)^2 +(\bar\psi{i}{\gamma}^{5}\vec\tau\psi)^2\right]
\mathrel{\mathop{\longrightarrow}^{\rm Fierz}}
{\cal L}_{E}^{(A)}
,
\ee
and defining a total Lagrangian
\be
{\cal L}^{(A)} = {\cal L}_{D}^{(A)} - {\cal L}_{E}^{(A)}
{}.
\label{eq:fierz}
\ee
The exchange terms of the original Lagrangian defined in eq.(\ref{eq:njla1})
are now included by evaluating the contributions of all terms in
eq.(\ref{eq:fierz}).
The second-order mass contribution for this Lagrangian is small
\cite{domi}
and dominated by the second-order self-energy for the pseudovector
interaction, while the second-order self-energy contributions
for the scalar and vector interactions essentially cancel.

Model B is defined by the Lagrangian:
\be
{\cal L}_{D}^{(B)}={\bar\psi}(i\nabla\hskip-0.8em{/}-m_0)\psi
-G_{B}(\bar\psi{\gamma}_{\mu}\psi)^2. \label{eq:njlb1}
\ee
The interaction term defined in this Lagrangian is symmetric
under a chiral transformation and is a scalar in color space. The
exchange terms are determined in a way analogous to
eq.(\ref{eq:fierz}). This Lagrangian yields
the largest second-order effect in which the real part of
the scalar self-energy is dominated by the coupling to vector
$q\bar{q}$ excitations.

The Lagrangian of model C:
\be
{\cal L}_{D}^{(C)}={\bar\psi}(i\nabla\hskip-0.8em{/}-m_0)\psi
-G_C(\bar\psi{\gamma}_{\mu}\vec\lambda\psi)^2,
\label{eq:njlc1}
\ee
may be of special interest as it can be motivated by QCD.
Starting from the path integral formulation of QCD, attempts have been
made to eliminate the gluon degrees of freedom by trying to derive
or motivate effective Lagrangians for quarks in the low-energy domain.
Employing a current expansion of the effective quark action
\cite{roberts}, or using the field strength approach \cite{schad,hugo},
one obtains a quark-quark interaction defined in terms of color-vector
currents. Using the nomenclature introduced before, the
corresponding NJL type Lagrangian exhibits the form of
eq.(\ref{eq:njlc1}). Here the second-order
mass contribution is also dominated by the vector interaction
for this Lagrangian.

Finally, Lagrangian D provides second-order mass contributions similar to that
of Lagrangian A, but significantly larger.
This Lagrangian has also been considered in ref. \cite{bueck}
, but with a momentum dependent coupling constant:
\be
{\cal L}_{D}^{(D)}={\bar\psi}(i\nabla\hskip-0.8em{/}-m_0)\psi
+G_{D}\left[(\bar\psi\vec\lambda\psi)^2
+(\bar\psi{i}{\gamma}^{5}\vec\tau\vec\lambda\psi)^2\right].
\label{eq:njld1}
\ee

Each of the Lagrangians considered are nonrenormalizable, due to the contact
interaction. Therefore one has to introduce a cutoff scheme to regularize
the various contributions.
In this work a three-momentum noncovariant cutoff scheme is utilized.
Hence, for each directed line in the second-order self-energy (Fig. 1),
which represents a single-particle propagator, the appropriate
three-momenta squared are restricted to values less than
the cutoff parameter $\Lambda$ squared, while the zeroth components
of the momenta are unrestricted.

Applying the same cutoff procedure to
the evaluation of the Hartree-Fock contribution to the self-energy, one
obtains a non-linear gap equation to determine the effective quark mass
in this approximation:
\be
m_{HF}^* = m_{0} + \tilde G\frac{24}{(2\pi )^3} \int_{0}^\Lambda d^3p \,
\frac{m_{HF}^*}{\sqrt{{\bf p}^2 + {m_{HF}^*}^2}} . \label{eq:gap}
\ee
The coupling constant $\tilde G$ refers to the coupling constant in
front of the scalar - isoscalar - colorscalar term of the Lagrangian
including the effects of the exchange term in eq.(\ref{eq:fierz}).
For the various models we obtain:
\be
\tilde G = \frac{13}{12}G_{A} = \frac{1}{6}G_{B} = \frac{8}{9}G_{C} =
\frac{4}{9}G_{D} . \label{eq:coupco}
\ee
Hence, different strengths for the residual interaction
in the various channels result when the coupling constants
are adjusted so that the Hartree-Fock approximation leads to
the very same gap equation for each model. As in our previous paper
\cite{domi} the coupling  constants $G_{i}$ for the various models
$i=A,B,C,D$ are adjusted in such a way that the effective mass
calculated from the Hartree-Fock plus the second order terms is
313 MeV for a quark of momentum $\vert \vec{p} \vert$ = 0. The cutoff
parameter $\Lambda$ = 653 MeV has been chosen for all models.

\section{Calculation of the Self-Energy}

The second-order contributions to the self-energy of the quarks can be
split into various parts $\Sigma^{(\lambda )}$, where the index
$\lambda$ refers to the quantum numbers of the $q\bar{q}$ excitation
considered. We distinguish $\lambda$ = $s$ for scalar, $\lambda$ = $ps$
for pseudoscalar, and $\lambda$ = $v$ for vector excitations. The
distinction of the $q\bar{q}$ excitations with respect to the flavor
and color quantum numbers will be discussed at the end of this section.

For a given excitation mode, $\lambda $, the scalar contribution
of the second-order self-energy is obtained as follows:
\be
\delta \Sigma^{( \lambda )}_s = \frac{1}{4n_{f}n_{c}}
\mbox{tr} \left[ \delta \Sigma ^{( \lambda )} \right] ,
\label{eq:defsc}
\ee
where $n_f$ and $n_c$ are the number of flavors and colors,
respectively. All traces in this paper are over spin (Dirac),
color, and flavor spaces.

In \cite{domi} it has been shown that the real part of these self-energy
contributions, $\delta \Sigma^{( \lambda )}_s$, can be evaluated from
irreducible response functions $\Pi^{(0)}_{\lambda}$ of the
corresponding excitation mode. The real and imaginary parts of
$\Pi^{(0)}_{\lambda}$ are related by a dispersion relation:
\be
\mbox{Re} \Pi^{(0)}_{\lambda} (q) =
\frac{1}{\pi} \mbox{P} \int_{0}^\infty dq_{0}' \,
\mbox{Im} \Pi^{(0)}_{\lambda}
(q_{0}', {\bf q} ) \left[ \frac{1}{q_{0}' - q_{0}}
+ \frac{1}{q_{0}' + q_{0}} \right] , \label{eq:disp}
\ee
and also the real and imaginary parts are even
functions of the energy variable $q_{0}$. Therefore the second-order
contributions to the scalar self-energy can be determined by knowing only
the imaginary part of the polarization propagator. As an example, we
recall the result for $\delta \Sigma^{( s )}_s$, the scalar
contribution to the self-energy originating from a pure scalar
interaction in a Lagrangian of the form:
\be
\delta {\cal L} = \tilde G (\bar\psi\psi)^2. \label{eq:scall}
\ee
Approximating the Green's functions needed for the evaluation of the
second-order diagram by those of a quark with a constant mass $m^*$, one
obtains:
\begin{eqnarray}
\lefteqn{\delta\Sigma_{s}^{(s)}(p) = -\frac{\tilde G m^*}{4 \pi^3}
\int_{0}^\Lambda \frac{{\bf q}^2  \, dq}{E^*_{q}} \int_{-1}^{+1} dx}
\nonumber \\
&&
\times
\biggl\lbrace  \int_{0}^\infty dq_{0} \frac{\mbox{Im}
\Pi^{(0)}_{s} (q_{0}, \vert {\bf p} - {\bf q} \vert )}
{q_{0}+p_{0}+E_{q}^*+i \epsilon}
+
\int_{0}^\infty dq_{0} \frac{\mbox{Im}
\Pi^{(0)}_{s} (q_{0}, \vert {\bf p} - {\bf q} \vert )}
{q_{0}-p_{0}+E_{q}^*-i \epsilon}
\biggr\rbrace.
\nonumber \\
\label{eq:sigs2}
\end{eqnarray}
The nomenclature used in this equation corresponds to the assignment of
momenta in Fig. 1. The integration variable $x$ represents the cosine
of the angle between ${\bf p}$ and ${\bf q}$. The imaginary part of the
scalar response function is:
\begin{eqnarray}
\mbox{Im} \Pi^{(0)}_{s} (q)& = &\frac{\tilde G n_{f} n_{c}}{4 \pi^2} \int
\frac{d^3k}{E^*_{k}E^*_{k-q}} \left( E^*_{k}E^*_{k-q} + {\bf k} ({\bf
k}- {\bf q}) - {m^*}^2 \right) \nonumber \\
&& \times \delta (\vert q_{0}\vert - E^*_{k} -
E^*_{k-q} ) \Theta (\Lambda - \vert {\bf k} \vert )
\Theta (\Lambda - \vert {\bf k } - {\bf q} \vert ) \; , \label{eq:pols}
\end{eqnarray}
with
\be
E^*_{k} = \sqrt{ {\bf k}^2 + {m^*}^2} .
\ee
This self-energy result is a simple extension of the real part found
previously \cite{domi}. Eq.(\ref{eq:sigs2}) furthermore demonstrates
that the imaginary part is:
\begin{eqnarray}
\delta \mbox{Im}\Sigma_{s}^{(s)}(p)&=&-\mbox{sgn}( p_0 )
\Theta ( p_0^2 - {m^*}^2 )
\frac{\tilde G m^*}{4 \pi^2} \nonumber \\ && \times
\int_{0}^{\mbox{min} ( \Lambda , \sqrt{p_0^2 - {m^*}^2} )}
\frac{{\bf q}^2  \, dq}{E^*_{q}} \int_{-1}^{+1} dx \, \mbox{Im}
\Pi^{(0)}_{s} ( \vert p_{0} \vert -E^*_{q}, \vert {\bf p} - {\bf q} \vert )
, \nonumber \\ \label{eq:imsigs2}
\end{eqnarray}
and also that the real and imaginary parts of the scalar contribution to
the self-energy are related by a dispersion relation:
\be
\delta \mbox{Re} \Sigma^{( s )}_{s} (q) =
\frac{1}{\pi} \mbox{P} \int_{0}^\infty dq_{0}' \,
\delta \mbox{Im} \Sigma^{( s )}_{s}
(q_{0}', {\bf q} ) \left[ \frac{1}{q_{0}' - q_{0}}
+ \frac{1}{q_{0}' + q_{0}} \right] . \label{eq:sedisp}
\ee
Hence, we find that the imaginary part $\delta
\mbox{Im}\Sigma_{s}^{(s)}(p)$ yields an odd function of $p_{0}$ for a
given $\vert {\bf p} \vert $ (see eq.(\ref{eq:imsigs2})), while the real
part is clearly even (see eq.(\ref{eq:sedisp})). As the expressions
obtained for the scalar contributions originating from pseudoscalar
($\lambda $ = $ps$) and vector excitation modes ($\lambda $ = $v$) are
similar to eq.(\ref{eq:sigs2}) (see \cite{domi}), the dispersion
relation of eq.(\ref{eq:sedisp}) is also valid for these modes. One
finds in general for all excitation modes $\lambda$ that, for a given
$\vert {\bf p} \vert $, the real part of $\delta \Sigma_{s}^{(\lambda)}(p)$
is identical for $p_{0}$ and $-p_{0}$, while the imaginary part changes
sign.

We now turn to the time-like vector contributions to the second-order
self-energy
originating from the various excitation modes $\lambda$. These
contributions are defined by:
\be
\delta \Sigma^{( \lambda )}_0 = \frac{-1}{4n_{f}n_{c}} \mbox{tr}
\left[ \gamma^0 \delta \Sigma^{( \lambda )} \right] ,
\ee
(compare eq.(\ref{eq:defsc})). Note that the Hartree-Fock approximation
does not provide a vector contribution to the self-energy for quarks
in the vacuum.
Each complex self-energy originating from the second-order terms is
derived in a manner similar to that
used to find the real part of the scalar self-energies
\cite{domi}.

In the case of scalar excitation modes ($\lambda$ = $s$), assuming a
scalar interaction as described by the Lagrangian
in eq.(\ref{eq:scall}), the result is:
\begin{eqnarray}
\delta\Sigma_{0}^{(s)}(p)&=& \frac{i \tilde G}{2} \int
\frac{d^4q}{(2\pi )^4} \mbox{tr} \left[ \gamma^0 g(q) \right]
\Pi^{(0)}_{s} (p -q )
\label{eq:sigs10}
\\
&=&
\frac{- \tilde G }{4 \pi^3}
\int_{0}^\Lambda {\bf q}^2 dq \int_{-1}^{+1} dx
\nonumber \\
&&
\times
\biggl\lbrace  \int_{0}^\infty dq_{0} \frac{\mbox{Im}
\Pi^{(0)}_{s} (q_{0}, \vert {\bf p} - {\bf q} \vert )}
{q_{0}+p_{0}+E_{q}^*+i \epsilon}
-
\int_{0}^\infty dq_{0} \frac{\mbox{Im}
\Pi^{(0)}_{s} (q_{0}, \vert {\bf p} - {\bf q} \vert )}
{q_{0}-p_{0}+E_{q}^*-i \epsilon}
\biggr\rbrace .
\nonumber \\
\label{eq:sigs20}
\end{eqnarray}
This result is quite similar to the corresponding result for the scalar
self-energy (see eq.(\ref{eq:sigs2})) except for the facts
that a factor $m^*/E_{q}^*$
is missing and a minus sign occurs in front of the second principal value
integral.
Taking the real part of eq.(\ref{eq:sigs20}), an odd function
of $p_0$ for fixed $\vert {\bf p} \vert$ is obtained.
The imaginary part of eq.(\ref{eq:sigs20}) is:
\begin{eqnarray}
\delta \mbox{Im}\Sigma_{0}^{(s)}(p)&=&
\Theta ( p_0^2 - {m^*}^2 )
\frac{\tilde G}{4 \pi^2}
\nonumber \\
&&
\times
\int_{0}^{\mbox{min} ( \Lambda , \sqrt{p_0^2 - {m^*}^2} )}
{{\bf q}^2  \, dq} \int_{-1}^{+1} dx
\,
\mbox{Im}
\Pi^{(0)}_{s} ( \vert p_{0} \vert -E^*_{q}, \vert {\bf p} - {\bf q} \vert )
. \nonumber \\
\label{eq:imsgs20}
\end{eqnarray}
The imaginary part clearly is an even function.

The vector second-order self-energy for a purely pseudoscalar interaction
is:
\be
\delta\Sigma_{0}^{(ps)}
(p)=\frac{-i \tilde G}{2} \int
\frac{d^4q}{(2\pi )^4} \mbox{tr} \left[ \gamma^0 g(q) \right]
\Pi^{(0)}_{ps} (p -q) . \label{eq:0pses1}
\ee
Comparing eq.(\ref{eq:sigs10}) and eq.(\ref{eq:0pses1}),
the complex self-energy and its imaginary part
in this case can be found by using
eq.(\ref{eq:sigs20}) and eq.(\ref{eq:imsgs20}), respectively.
For the vector self-energy with a pure scalar interaction, a
positive value is obtained for energies less than the cutoff,
and the vector self-energy with a pseudoscalar interaction provides
a negative result in the same energy domain.

The vector self-energy for the case of a pure vector interaction
is more easily calculated by using different momenta than those shown
in Fig. 1. For this case, $p$ and $p-q$ are simply exchanged.
Then,
\begin{eqnarray}
\delta\Sigma_{0}^{(v)}
(p)&=& \frac{-i\tilde G}{4} \int
\frac{d^4q}{(2\pi )^4}
\sum_{\theta ,\phi}
\mbox{tr}\left[
\gamma^0 \gamma_{\theta} g(p - q) \gamma^{\phi}\right]
\Pi^{(0) \theta}_{v , \phi} (q)
\\
\label{eq:0vses1}
&=&\frac{\tilde G}{4 \pi^3}
\int_{-1}^{+1} dx
\int_{0}^{{\vert {\bf p} \vert} x + \sqrt{{\Lambda}^2
- {\vert {\bf p} \vert}^2 ( 1 - x^2 )}} {{\bf q}^2  \, dq}
\nonumber \\
&&
\times
\biggl\lbrace{  \int_{0}^\infty dq_{0}
\frac{\mbox{Im}
\,
\mbox{f}^{(0)}_{v}
(q_{0},\vert {\bf p} \vert,\vert {\bf q} \vert,x)}
{q_{0}+p_{0}+E_{p-q}^*+i \epsilon}
-
\int_{0}^\infty dq_{0}
\frac{\mbox{Im}
\,
\mbox{f}^{(0)}_{v}
(q_{0},\vert {\bf p} \vert,\vert {\bf q} \vert,x)}
{q_{0}-p_{0}+E_{p-q}^*-i \epsilon}
\biggr\rbrace} ,
\nonumber \\
\label{eq:0sigv2}
\end{eqnarray}
for $\vert {\bf p} \vert < \Lambda $.
For the case of $\vert {\bf p} \vert \geq \Lambda $,
the lower limit of the integral over $x$ is replaced by
$\sqrt{1 - ({\frac{\Lambda}{\vert {\bf p} \vert}})^2}$.
The imaginary part of this self-energy is:
\be
\delta \mbox{Im}\Sigma_{0}^{(v)}(p)=
- \Theta ( p_0^2 - {m^*}^2 )
\frac{\tilde G }{4 \pi^2}
\int_{-1}^{+1} dx
\int_{0}^{q_{max}}{{\bf q}^2  \, dq}
\,
\mbox{Im}
\,
\mbox{f}^{(0)}_{v}
( \vert p_{0} \vert -E^*_{q}, \vert {\bf p} \vert, \vert {\bf q} \vert,x ) ,
\label{eq:imsgv20}
\ee
for $\vert p \vert < \mbox{min} ( {\Lambda}^2 , {p_0}^2 - {m^*}^2 ) $,
and
\be
q_{max}=
\vert {\bf p} \vert x
+ \mbox{min} \left[
{\sqrt{{\Lambda}^2 - {\vert {\bf p} \vert}^2 ( 1 - x^2 )}} ,
{\sqrt{({p_0}^2 - {m^*}^2)
- {\vert {\bf p} \vert}^2 ( 1 - x^2 )}} \, \right] .
\ee
For $\vert {\bf p} \vert \geq \mbox{min} ( {\Lambda}^2 , {p_0}^2 - {m^*}^2 ) $,
the lower limit of the $x$ integral is changed in a manner
similar to that done in eq.(\ref {eq:0sigv2})
when $\vert {\bf p} \vert \geq \Lambda $.

The imaginary part of the function f in this self-energy
is given by:
\begin{eqnarray}
\mbox{Im}
\,
\mbox{f}^{(0)}_{v}
( q_{0} , \vert {\bf p} \vert, \vert {\bf q} \vert,x )& = &
\mbox{Im}\Pi^{(0) 0}_{v , 0}
(q_{0},\vert {\bf q} \vert )
\nonumber
\\
&&
\mbox{ } - \sum^3_{i=1} \left[
\mbox{Im}\Pi^{(0) i}_{v , i}
(q_{0},\vert {\bf q} \vert )
-
\frac{2}{E^*_{p-q}}
(\mbox{p}_{i} - \mbox{q}_{i})
\mbox{Im}\Pi^{(0) 0}_{v , i}
(q_{0},\vert {\bf q} \vert )
\right]
,
\nonumber
\\
\label{eq:imf}
\end{eqnarray}
and behaves as the response functions encountered here.
The various components of the matrix for the vector response
function are obtained from the following formula:
\begin{eqnarray}
\Pi^{(0) {\theta}}_{v,\phi} (q) =
-2i\tilde G \int \frac{d^4k}{(2\pi )^4} \mbox{tr}
\left[ \gamma^{\theta} g(k)\gamma_{\phi} g(k-q) \right]
{}.
\end{eqnarray}
The terms of the function $\mbox{f}$ in eq.(\ref{eq:imf}) can
be separated and written as:
\begin{eqnarray}
\mbox{Im}\Pi^{(0) 0}_{v , 0}
(q_{0},\vert {\bf q} \vert )
&-& \sum^3_{i=1}
\mbox{Im}\Pi^{(0) i}_{v , i}
(q_{0},\vert {\bf q} \vert )
 =
-\frac{\tilde G n_{f} n_{c}}{2 \pi^2} \int
\frac{d^3k}{E^*_{k}E^*_{k-q}} \left( 2E^*_{k}E^*_{k-q} +
{m^*}^2 \right) \nonumber \\
&& \times \delta (\vert q_{0}\vert - E^*_{k} -
E^*_{k-q} ) \Theta (\Lambda - \vert {\bf k} \vert )
\Theta (\Lambda - \vert {\bf k } - {\bf q} \vert ) \; , \label{eq:polvo1}
\end{eqnarray}
and
\begin{eqnarray}
\sum^3_{i=1}
(\mbox{p}_{i} - \mbox{q}_{i})
\mbox{Im}\Pi^{(0) 0}_{v , i}
(q_{0},\vert {\bf q} \vert )
& = &
\mbox{sgn}( q_0 )\frac{\tilde G n_{f} n_{c}}{4 \pi^2} \int
\frac{d^3k}{E^*_{k}E^*_{k-q}}
\nonumber
\\
&&
\times
\left[
({\bf p} - {\bf q}) {\bf k}
(E^*_{k-q}-E^*_{k}) +
({\bf p} - {\bf q}) {\bf q} E^*_{k} \right] \nonumber \\
&& \times \delta (\vert q_{0}\vert - E^*_{k} -
E^*_{k-q} ) \Theta (\Lambda - \vert {\bf k} \vert )
\Theta (\Lambda - \vert {\bf k } - {\bf q} \vert ) \; .
\nonumber
\\
\label{eq:polvo2}
\end{eqnarray}

{}From eqs.(\ref{eq:imsgs20}) and (\ref{eq:imsgv20}) one can see that, for
all excitation modes $\lambda$, the imaginary parts of the vector
contributions are even functions of $p_{0}$, while eqs.(\ref{eq:sigs20})
and (\ref{eq:0sigv2}) demonstrate that the real parts are odd functions of
$p_{0}$. The real and imaginary parts of the second-order vector
self-energies are related by:
\be
\delta \mbox{Re} \Sigma^{( \lambda )}_{0} (q) =
\frac{1}{\pi} \mbox{P} \int_{0}^\infty dq_{0}' \,
\delta \mbox{Im} \Sigma^{( \lambda )}_{0}
(q_{0}', {\bf q} ) \left[ \frac{1}{q_{0}' - q_{0}}
- \frac{1}{q_{0}' + q_{0}} \right] . \label{eq:sedispo}
\ee

Until now we have only considered the Dirac structure of the excitation
modes $\lambda$ and the corresponding interaction terms, assuming a
scalar for the isospin and color part. For interaction terms
which are isovector rather than isoscalar,
one obtains the corresponding second-order contribution to the
self-energy with an additional factor of 3.
Also, for interaction terms which are products of vector
operators in SU(3) color space,
the corresponding contributions to the self-energy should be multiplied
by a factor of 32/9.

\section{Results and Discussion}

As a first point in the discussion of our results, we would like to
consider several of the basic features of the various
contributions to the quark
self-energy. For that purpose we intially consider
in our discussion highly simplified quark-quark interactions which
contain either a pure scalar contribution (see eq.(\ref{eq:scall})),
exciting only scalar $q\bar{q}$ excitations, or similarly
pure pseudoscalar or vector
contributions, while always ignoring the Fock exchange terms. To allow a
comparison between these different types of self-energies, we have
chosen the same
coupling constant and cutoff parameter for these modes ($\tilde G$ =
1.618 $\times 10^{-6}$ MeV$^{-2}$,
$\Lambda$ = 653 MeV, see \cite{domi}). We would like to recall that, in
calculating the contributions of second-order, we consider a Green's
function for quarks with a constant effective mass $m^*$ of 313 MeV.

Results for the imaginary parts of the scalar and vector self-energies
originating from a pure scalar interaction are displayed in Fig. 2.
The various momenta ($\vert {\bf p}\vert$ = 0, 300, and 600 MeV) are
considered for the quarks. It is clear that the imaginary part can be
different from zero only for those positive energies $p_{0}$ ranging
between the threshold of $2q-1\bar{q}$ excitations
($3m^*$ = 939 MeV) and the
largest energy compatible with the cutoff ($3\sqrt{{m^*}^2 + \Lambda^2}
\approx$ 2170 MeV). As can be deduced from eqs.(\ref{eq:imsigs2}) and
(\ref{eq:imsgs20}), the imaginary part of the scalar self-energy is negative,
while the imaginary part of the vector self-energy is positive. Note
that, for negative energies $p_{0}$, the imaginary part of the scalar
self-energy changes sign, whereas the imaginary part of vector
self-energy depends on $\vert p_{0} \vert $ only (see discussion in
Sec. 3). The absolute values of the imaginary parts decrease with an
increase of the space-like component $\vert {\bf p} \vert$ of the quark
momentum. This leads to a reduction by almost a factor of 2, if the
momentum is increased from $\vert p_{0} \vert $ = 0 to $\vert p_{0} \vert $
= 600 MeV. A quite similar momentum dependence of the imaginary
contributions to the self-energy is also observed if other interactions
are considered.

Therefore, for the comparison of contributions arising from the three
basic kinds of interactions, displayed in Figure 3, results are
presented only for $\vert \bf{p} \vert $ = 0. It is worth noting that
the contributions originating from scalar and pseudoscalar interactions
always have opposite sign. Since the coupling strengths of these two
kinds of interactions are connected in Lagrangians with chiral symmetry
(see, e.g.~, A and D, defined in eqs.(\ref{eq:njla1}) and
(\ref{eq:njld1})), these contributions tend to cancel each other. Large
contributions are obtained from vector interactions and corresponding
$q\bar{q}$ excitations.

As discussed in Section 3, the real and imaginary parts
of the second-order
contributions to the self-energy are related by the dispersion
relations of eq.(\ref{eq:sedisp}) and eq.(\ref{eq:sedispo}) for the
scalar and vector terms, respectively. From these dispersion relations
one can immediately understand the features of the energy-dependence of
the real parts of the self-energies shown in Fig. 4, which are related
to the corresponding imaginary parts displayed in Fig. 3. It is
obvious from eq.(\ref{eq:sedispo}) that the real vector part of the
self-energy is an odd function of the energy variable $p_{0}$ and
consequently vanishes for $p_{0}=0$, while the real scalar part is an
even function and therefore can be different from 0 also at $p_{0}=0$.
With increasing positive energies, the real parts
of both the scalar and vector self-energies become larger in absolute
value with increasing $p_{0}$ until one approaches energies around
1500 MeV, i.e.~, energies close to the maximum of the imaginary parts. At
these energies the real parts change sign and finally approach zero.

In discussing the mathematical behavior of the real part, one should keep
in mind, however, that the maximum of the absolute value of the
imaginary part of the second-order
self-energy is primarily determined by the cutoff procedure for
the intermediate $2q-1\bar{q}$ excitations (see discussion above).
Therefore the change in sign of the real part of the self-energy,
which is related to this maximum, should not seriously be considered
from a physical point of view. Indeed, as we will see below, the
on-shell solution of the Dirac equation for quarks with momenta below
the cutoff ($\vert {\bf p} \vert < \Lambda$) requires self-energies
at energies $p_{0}$ below those for which the real part changes sign.

The statements on the cancellation between contributions arising from
scalar and pseudoscalar interactions made above for the imaginary
part, of course, also apply to the real parts. From the dispersion
relations it is furthermore obvious that the various real parts
depend on the space-component ${\bf p}$ of the quark
momentum in a similar way as the corresponding imaginary parts.

Having discussed several of the basic features of the various
contributions to the self-energy arising from the different interaction
terms, we now turn to the 4 effective Lagrangians defined as models A,
B, C and D in Section 2. For each of these Lagrangians, we calculate the
contributions to the quark self-energy of first and second order in the quark
- quark interaction and consider the on-shell solutions of the Dirac
equation for quarks with this self-energy. For each momentum $\vert
{\bf p} \vert$, these on-shell solutions must fulfill the
following energy relation:
\begin{eqnarray}
\left(p_{0} +  \mbox{Re}\Sigma_{0}^{(2)}(p_{0},{\bf p}) \right)^2
& = & {\bf p}^2 + \left( m_{0} + \mbox{Re}\Sigma_{s}^{HF} +
\mbox{Re}\Sigma_{s}^{(2)}(p_{0},{\bf p}) \right)^2 ,
\label{eq:0meff}
\end{eqnarray}
where $\Sigma_{0}^{(2)}$ and $\Sigma_{s}^{(2)}$ refer to the vector
and scalar components, respectively, obtained from the second-order terms.
For each
Lagrangian we assume a bare current quark mass of $m$ = 5 MeV and
adjust the coupling constant in such a way that the relation of
eq.(\ref{eq:0meff}) yields an energy $p_{0}$ of 313 MeV for a quark
with momentum $\vert {\bf p} \vert$ = 0. In order to get an idea of the
importance of the various contributions to the self-energy in the
different models, we present in Table 1 these on-shell quantities for
quarks with zero momentum $\vert {\bf p} \vert$.

One finds that the relative importance of the second-order contributions
as compared to the Hartree-Fock term is quite different in the
different Lagrangians under consideration. The effect of the
second-order terms is negligibly small for model A, which corresponds
to the original NJL Lagrangian (see eq.(\ref{eq:njla1})). This is in
accord with the observation made above that effects due to a scalar and a
pseudoscalar interaction, which are the dominating components for this
model, tend to cancel in the second-order contributions to the
self-energy. The other
extreme is model B (see eq.(\ref{eq:njlb1})), for which the second-order
scalar term is even larger than the Hartree-Fock contribution to the
scalar term. Since, however, the contributions of the second-order
scalar and vector terms to the energy $p_{0}$ partly cancel each other,
the total effect of the second-order terms on $p_{0}$ is smaller than
the Hartree-Fock term. Such a partial cancellation also occurs for the
Lagrangian of model C (see eq.(\ref{eq:njlc1})), which is
motivated from QCD. It should be kept in mind, however, that, in
this case also, individual contributions are as large as 15 percent of the
leading Hartree-Fock term. Substantial second-order effects are also
observed in model D.

While the Hartree-Fock contribution to the self-energy yields a
constant, the second-order contributions depend on the four-momentum
$p$ of the quark under consideration. This implies that the
quasiparticle energy, i.e., the self-consistent solution for $p_{0}$,
$p_{0}^{QP} (\vert {\bf p}\vert)$,
obtained from eq.(\ref{eq:0meff}), is different from the quasiparticle
energy for a free Dirac particle of mass $m^*$ or the
quasiparticle energy resulting from a Hartree-Fock approximation, if the
coupling constant is chosen to yield an effective mass $m^*$. To
visualize the effects of the momentum-dependent self-energy on the
quasiparticle energy $p_{0}^{QP} (\vert {\bf p}\vert)$ we define an
enhancement factor:
\be
{\cal E} (\vert {\bf p}\vert) = \frac{p_{0}^{QP} (\vert {\bf
p}\vert)}{\sqrt{\vert {\bf p}\vert^2 + {m^*}^2}} .
\label{eq:enhanc}
\ee
This enhancement factor is displayed in the upper part of Fig. 6. As
the coupling constants for all four Lagrangians were adjusted to yield
a quasiparticle energy $p_{0}^{QP}$ = $m^*$ for zero momentum, it is
clear that all enhancement functions approach the value 1 in this
limit. For larger momenta, however, the enhancement or better reduction
factor deviates from this value indicating that the dependence
of the self-energy on energy and momentum leads to quasiparticle
energies which are
quite different from those of a free particle.

It is worth noting that
the larger fraction of this deviation is due to the vector component in
the self-energy in models B and C. This can be seen by comparing
the enhancement factors
obtained in our complete calculation (upper part of the figure) to
enhancement factors which are obtained if only the scalar part of the
second-order self energy is taken into account (see the lower
part of the figure, where the corresponding enhancement factors are
displayed, resulting from \cite{domi}).
Also one may recall that
the energy-dependence of the vector self-energy, which is an odd
function of $p_{0}$, is generally larger than the corresponding scalar
contribution, which is an even function of $p_{0}$ (see figure 4).
{}From the enhancement factors displayed in Fig. 6, it is obvious that
Lagrangians dominated by vector quark-quark interactions (models B and
C) tend to show much larger effects due to the second-order vector
contributions
than those which contain direct interaction terms of scalar and
pseudoscalar form.

\section{Conclusions}
The self-energy of quarks is calculated for various models which are
inspired by the Nambu--Jona-Lasinio (NJL) model, including terms up to
second-order in the self-energy of the quarks. The real and imaginary parts
of scalar and vector components in the self-energy are discussed. It
is demonstrated that the effects of the second-order terms are
particularly large if the Lagrangian contains a strong vector
quark-quark interaction. In such a model (Table 1, model B) individual
contributions to the real part of the self-energy can even be larger
than the Hartree-Fock contribution in the same model.

While the self-energy calculated in the Hartree-Fock approximation
yields a constant, the contributions of second order display a
characteristic dependence on the energy and momentum of the quark under
consideration. The solution of the Dirac equations for quarks with such
a self-energy leads to relations between the energy and momentum of a
quark which are different from that of a free Dirac particle with a constant
effective mass. At small energies, it is the energy dependence of the
vector component of the self-energy, in particular, which yields
important modifications.

The second-order self-energies are directly related to the bare
$q\bar{q}$ polarization propagators $\Pi^{(0)}_{\lambda}$ in the
different excitation modes $\lambda$. If, in a next step, one replaces
the bare propagators by the reducible polarization propagators as
obtained in the RPA approximation, one expects a non-vanishing
imaginary part in the self-energy at energies below the $2q-1\bar{q}$
threshold displayed, e.g., in Fig. 3. As the real and imaginary parts
of the
self-energy are related by dispersion relations (see
eqs.(\ref{eq:sedisp}) and (\ref{eq:sedispo})), this implies an even
stronger energy dependence than that obtained in the approach presented
here. Investigations in this direction are in progress.

This work has partly been supported by the Deutsche
Forschungsgemeinschaft (DFG Fa 67/14-1); this support is gratefully
acknowledged.

%        bibliography

%\end{references}
\clearpage
\begin{table}[h]
\caption{Various contributions to the real part of the
self-energy for quarks with momenta ${\bf p}$ = 0. The row labeled
$\Sigma_{s}^{HF}$ shows the Hartree-Fock contribution, while the labels
$\Sigma_{s}^{(2)}$ and $\Sigma_{0}^{(2)}$ refer to the scalar and
vector terms, respectively, of the real part of the self-energy obtained in
second-order. Results
from the various Lagrangians A, B, C, and D, discussed in section 2,
are compared. For each Lagrangian the coupling constant has been chosen
such that the self-consistent energy relation of eq.(31) yields 313
MeV. All entries are given in MeV.}
\begin{center}
\begin{tabular}{crrrr}
&&&&\\ \hline\hline
&&&&\\
\multicolumn{1}{c}{Term}&\multicolumn{1}{c}{Model A}&
\multicolumn{1}{c}{Model B}&\multicolumn{1}{c}{Model C}&
\multicolumn{1}{c}{Model D}\\
&&&&\\ \hline
&&&&\\
$\Sigma_{s}^{HF}$   &  295.45  &   187.36 &    280.32 &       209.04 \\
$\Sigma_{s}^{(2)}$ &  10.67   &   203.60 &    44.97  &     87.25   \\
$\Sigma_{0}^{(2)}$  &  -1.88   &    82.96 &    17.29  &     -11.71    \\
&&&&\\ \hline\hline
\end{tabular}
\end{center}
\end{table}

\clearpage
\section {Figure Captions}
\bigskip\bigskip\noindent{\bf Figure 1:}
{Diagrams representing the second-order contribution to the self-energy
for a Fermion with momentum $p$. The labels for the momenta of the
intermediate Fermion ($q$) and the polarization and boson propagators
($p-q$) are identical to those used in Section 3.}

\bigskip\bigskip
\noindent{\bf Figure 2:}
{Imaginary parts of the scalar (left side) and vector (right side)
self-energies for quarks with momenta $p$, assuming a scalar interaction
(eq.(11)). For various values of $\vert{\bf p}\vert$, results are
displayed as a
function of $p_{0}$. The self-energy has been calculated assuming a
Green's function for quarks characterized by a constant effective
mass $m^*$ = 313 MeV, a cutoff parameter $\Lambda$ = 653 MeV,
and a coupling constant $\tilde G$ = 1.618 $\times 10^{-6}$
MeV$^{-2}$.}

\bigskip\bigskip\noindent{\bf Figure 3:}
{Imaginary parts of the scalar (left side) and vector (right side)
self-energies for quarks obtained from scalar, pseudoscalar, and vector
quark-quark interactions. Results are given for quarks with momentum
$\vert{\bf p}\vert =0$ as a function of energy $p_{0}$. While the
results for the vector component obtained from a vector interaction (line
labeled with asterisks in the right part of the figure) can be read with
the scale on the left-hand side, all other curves are related to the
scale on the right-hand side of the figure. For further details see Figure
2.}

\bigskip\bigskip\noindent{\bf Figure 4:}
{Real parts of the scalar and vector self-energies for quarks with
momentum $\vert{\bf p}\vert$ =0, assuming various interactions.
For further details, see the caption of Figure 3.}

\bigskip\bigskip\noindent{\bf Figure 5:}
{Real parts of the scalar and vector self-energies for quarks with
momentum $p$, considering various interactions. Assuming $p_{0}$ = 1
GeV, results are presented as a function of $\vert{\bf p}\vert$.
For further details, see the caption of Figure 3.}

\bigskip\bigskip\noindent{\bf Figure 6:}
{Enhancement factor ${\cal E}$ for the quasiparticle energies as
defined in eq.(32) as a function of the quark momentum ${\bf p}$ for
the various Lagrangians discussed in Section 2. While the upper part of
the figure displays the results obtained with inclusion of all
second-order contributions, the results displayed in the lower part are
obtained for the case when the vector component of the self-energy
is ignored.}


\begin{thebibliography}{99}
\bibitem{njl} Y. Nambu and G. Jona-Lasinio, Phys. Rev. {\bf 122} 345 (1961)
; {\bf 124} 246 (1961).
\bibitem{klev} S.P. Klevansky, Rev. Mod. Phys. {\bf 64}, 649 (1992).
\bibitem{alkof} R. Alkofer and H. Reinhardt, Z. Phys. {\bf A343} 79 (1992).
\bibitem{ulf} V. Bernard, U.-G. Meissner, and I. Zahed, Phys. Rev. {\bf
D36}, 819 (1987).
\bibitem{hats} T. Hatsuda and T. Kunihiro, Phys. Lett. {\bf B 185}, 304
(1987).
\bibitem{hen} E.M. Henley and H. M\"uther, Nuc. Phys. {\bf A513} 667 (1990).
\bibitem{cao} Nan-Wei Cao, C.M. Shakin, and Wei-Dong Sun, Phys. Rev.
{\bf C46} 2535 (1992).
\bibitem{domi} P.P. Domitrovich, D. B\"uckers, and H. M\"uther, Phys. Rev.
{\bf {C48}} 413 (1993).
\bibitem{serot} B.D. Serot and J.D. Walecka, Adv. Nucl. Phys. {\bf 16},
1 (1986).
\bibitem{chin} S.A. Chin, Ann. of Phys. {\bf 108}, 301 (1977).
\bibitem{roberts} R.T. Cahill and C.D. Roberts, Phys. Rev. {\bf D36}
2419 (1985).
\bibitem{schad} M. Schaden, H. Reinhardt, P.A. Amundsen, and M.J.
Lavelle, Nucl. Phys. {\bf B339}, 595 (1990).
\bibitem{hugo} H. Reinhardt, Phys. Lett. {\bf B248} 365 (1990).
\bibitem{bueck} D. B\"uckers and H. M\"uther, Nucl. Phys. {\bf A523}
629 (1991).

\end{thebibliography}
\end{document}